# A MULTIVARIATE SEMIPARAMETRIC BAYESIAN SPATIAL MODELING FRAMEWORK FOR HURRICANE SURFACE WIND FIELDS

By Brian J. Reich[1] and Montserrat Fuentes[2]

*North Carolina State University*

Storm surge, the onshore rush of sea water caused by the high winds and low pressure associated with a hurricane, can compound the effects of inland flooding caused by rainfall, leading to loss of property and loss of life for residents of coastal areas. Numerical ocean models are essential for creating storm surge forecasts for coastal areas. These models are driven primarily by the surface wind forcings. Currently, the gridded wind fields used by ocean models are specified by deterministic formulas that are based on the central pressure and location of the storm center. While these equations incorporate important physical knowledge about the structure of hurricane surface wind fields, they cannot always capture the asymmetric and dynamic nature of a hurricane. A new Bayesian multivariate spatial statistical modeling framework is introduced combining data with physical knowledge about the wind fields to improve the estimation of the wind vectors. Many spatial models assume the data follow a Gaussian distribution. However, this may be overly-restrictive for wind fields data which often display erratic behavior, such as sudden changes in time or space. In this paper we develop a semiparametric multivariate spatial model for these data. Our model builds on the stick-breaking prior, which is frequently used in Bayesian modeling to capture uncertainty in the parametric form of an outcome. The stick-breaking prior is extended to the spatial setting by assigning each location a different, unknown distribution, and smoothing the distributions in space with a series of kernel functions. This semiparametric spatial model is shown to improve prediction compared to usual Bayesian Kriging methods for the wind field of Hurricane Ivan.

Received November 2006; revised March 2007.
[1]Supported by NSF Grant DMS-03-54189.
[2]Supported in part by NSF Grant DMS-03-53029.
Supplementary material available at http://imstat.org/aoas/supplements

**1. Introduction.** Modeling surface wind fields is essential for hurricane forecasting. A wind field gives the wind velocity at any location in the vicinity of the hurricane. The numerical ocean models used to predict the storm surge for coastal areas rely heavily on wind field inputs. Currently, deterministic formulas such as the Holland model [Holland (1980)] are used to generate the wind fields for the storm surge model based on a few meteorological inputs such as the radius and central pressure of the storm.

While the Holland model captures many of the important features of a wind field, Foley and Fuentes (2006) show that this model does not allow for asymmetries often seen in wind fields and that storm surge prediction can be improved by supplementing the Holland model with a Gaussian geostatistical model. Another approach would be to introduce a more sophisticated deterministic wind model. A coupled atmospheric–oceanic numerical model can be used to simulate the surface winds at the boundary layer of the ocean model. However, the CPU time required to produce these modeled winds at high enough resolution for coastal prediction (1 to 4 km grids) prevents such model runs from being used in real-time applications. Alternatively, one could write a stochastic version of the deterministic model and approximate the physical model using a stochastic spatial basis. This is the approach of Wikle et al. (2001) for oceanic surface winds.

This paper proposes a semiparametric multivariate spatial model to predict a wind field. The predictions in this paper are purely spatial predictions made using multiple sources of observed data and Holland model output from a single time point. Several Gaussian multivariate spatial covariance models have been proposed. For example, Brown, Le and Zidek (1994) model the joint covariance of the observed multivariate data using an inverse Wishart distribution centered on a separable covariance matrix. Another approach is to represent the multivariate spatial process as a linear combination of univariate spatial process. Variations of this linear model of coregionization (LMC) have been used by Grzebyk and Wackernagel (1994), Wackernagel (2003), Schmidt and Gelfand (2003), Banerjee, Carlin and Gelfand (2004) and Gelfand et al. (2004). Foley and Fuentes (2006) apply the LMC to the two orthogonal west/east and north/south components of hurricane wind vectors.

Spatial models often assume the outcomes follow normal distributions. The Gaussian assumption is difficult to verify empirically and may be overly-restrictive for hurricane wind field data, which can display erratic behavior, such as sudden changes in time or space. For example, on the periphery of the map in Figure 1(a) the wind vectors vary smoothly from one measurement to the next. However, near the eye of the hurricane (center of the plot), the wind vectors are extremely volatile. Traditional Gaussian spatial models tend to oversmooth the area near the eye of the hurricane, resulting in a poor fit. Therefore, in this paper we develop a new multivariate semiparametric spatial model for these data that avoids specifying a Gaussian distribution for the spatial random effects.

Our semiparametric model avoids assuming normality by extending the stick-breaking prior of Sethuraman (1994) to the multivariate spatial setting. For general (nonspatial) Bayesian modeling, the stick-breaking prior offers a way to model a distribution of a parameter as an unknown quantity to be estimated from the data. The stick-breaking prior for the unknown distribution $F$ is the mixture

$$F \stackrel{d}{=} \sum_{i=1}^{m} p_i \delta(\theta_i), \tag{1}$$

where the number of mixture components $m$ may be infinite, $p_i$ are the mixture probabilities, and $\delta(\theta_i)$ is the Dirac distribution with point mass at $\theta_i$. The mixture probabilities "break the stick" into $m$ pieces so the sum of the pieces is one, that is, $\sum_{i=1}^{m} p_i = 1$. The first mixture probability is modeled as $p_1 = V_1$, where $V_1 \sim \text{Beta}(a, b)$. Subsequent mixture probabilities are $p_i = (1 - \sum_{j=1}^{i-1} p_j)V_i$, where $1 - \sum_{j=1}^{i-1} p_j$ is the probability not accounted for by the first $i-1$ mixture components, and $V_i \stackrel{\text{i.i.d.}}{\sim} \text{Beta}(a, b)$ is the proportion of the remaining probability assigned to the $i$th component. The locations $\theta_i \stackrel{\text{i.i.d.}}{\sim} F_o$, where $F_o$ is a known prior distribution. A special case of this prior is the Dirichlet process prior with $m = \infty$ and $a = 1$ [Ferguson (1973, 1974)].

The stick-breaking prior in (1) has been extended to the univariate spatial setting by incorporating spatial information into either the model for the locations $\theta_i$ or the model for the masses $p_i$. Gelfand, Kottas and MacEachern (2005a) and Gelfand, Guindani and Petrone (2007) model the locations as vectors drawn from a spatial distribution. This approach is generalized by Duan, Guindani and Gelfand (2007) to allow both the weights and locations to vary spatially. However, these approaches require replication, and thus are not appropriate for analyzing the wind fields data. Griffin and Steel (2006) propose a spatial Dirichlet model that does not require replication. Their model permutes the $V_i$ based on spatial location, allowing the prior to be different in different regions of the spatial domain.

This paper is the first to extend the stick-breaking prior to the multivariate spatial setting. Our semiparametric multivariate spatial model for a hurricane wind field has bivariate normal priors for the locations $\boldsymbol{\theta}_i$. Similar to Griffin and Steel, the probabilities $p_i$ vary spatially. However, rather than a random permutation of $V_i$, we introduce a series of kernel functions to allow the masses to change with space. This results in a flexible spatial model, as different kernel functions lead to different relationships between the distributions at nearby locations. This model is similar to that of Dunson and Park (2007), who use kernels to smooth the weights in the non-spatial setting. Our model is also computationally convenient because it avoids reversible jump MCMC steps and inverting large matrices which is crucial for analysis of hurricane wind fields since estimates must be made in real time.

The paper proceeds as follows. Section 2 describes the various sources of data used to map the wind field. The semiparametric spatial prior for univariate spatial data is introduced in Section 3. This model is extended to a multivariate model to analyze wind field data in Section 4. The model incorporates both a deterministic wind model and multiple sources of wind observations and allows for potential bias for each data source. This model is used to map the wind field of Hurricane Ivan in Section 5. Section 6 concludes.

**2. Description of the wind fields data.** We model wind fields data from Hurricane Ivan as it passed through the Gulf of Mexico at 12 pm on September 15, 2004. The three sources of information used in this analysis are plotted in Figure 1. The first source is gridded satellite data [Figure 1(a)] available from NASAs SeaWinds database (http://podaac.jpl.nasa.gov/products/product109.html). These data are available twice daily on a $0.25 \times 0.25$ degree global grid. Due to the satellite data's potential bias, measurement error and course temporal resolution, we supplement our wind fields analysis with data from NOAA's National Data Buoy Center. Buoy data are collected every ten minutes at a relatively small number of marine locations [Figure 1(b)]. These measurements are adjusted to a common height of 10 meters above sea level using the algorithm of Large and Pond (1981).

In addition to satellite and buoy data, our model incorporates the deterministic Holland model [Holland (1980)]. The NOAA currently uses this model alone to produce wind fields for their numerical ocean models. The Holland model predicts that the wind speed at location $\mathbf{s}$ is

$$H(\mathbf{s}) = \left(\frac{B}{\rho}\left(\frac{Rmax}{r}\right)^{B}(P_n - P_c)\exp\left[-\left(\frac{Rmax}{r}\right)^{B}\right]\right)^{1/2}, \tag{2}$$

where $r$ is the radius (km) from the storm center to site $\mathbf{s}$, $P_n$ is the ambient pressure (mb), $P_c$ is the hurricane central pressure (mb), $\rho$ is the air density (kg m$^{-3}$), $Rmax$ is radius of the maximum wind (km), and $B$ controls the shape of the pressure profile.

Section 4's multivariate spatial model decomposes the wind vectors into their orthogonal west/east ($u$) and north/south ($v$) vectors. The Holland model for the $u$ and $v$ components is

$$H_u(\mathbf{s}) = H(\mathbf{s})\sin(\phi) \quad \text{and} \quad H_v(\mathbf{s}) = H(\mathbf{s})\cos(\phi), \tag{3}$$

where $\phi$ is the inflow angle at site $\mathbf{s}$, across circular isobars toward the storm center, rotated to adjust for the storm's direction. We fix the parameters $P_n = 1010$ mb, $P_c = 939$ mb, $\rho = 1.2$ kg m$^{-3}$, and $Rmax = 49$ and

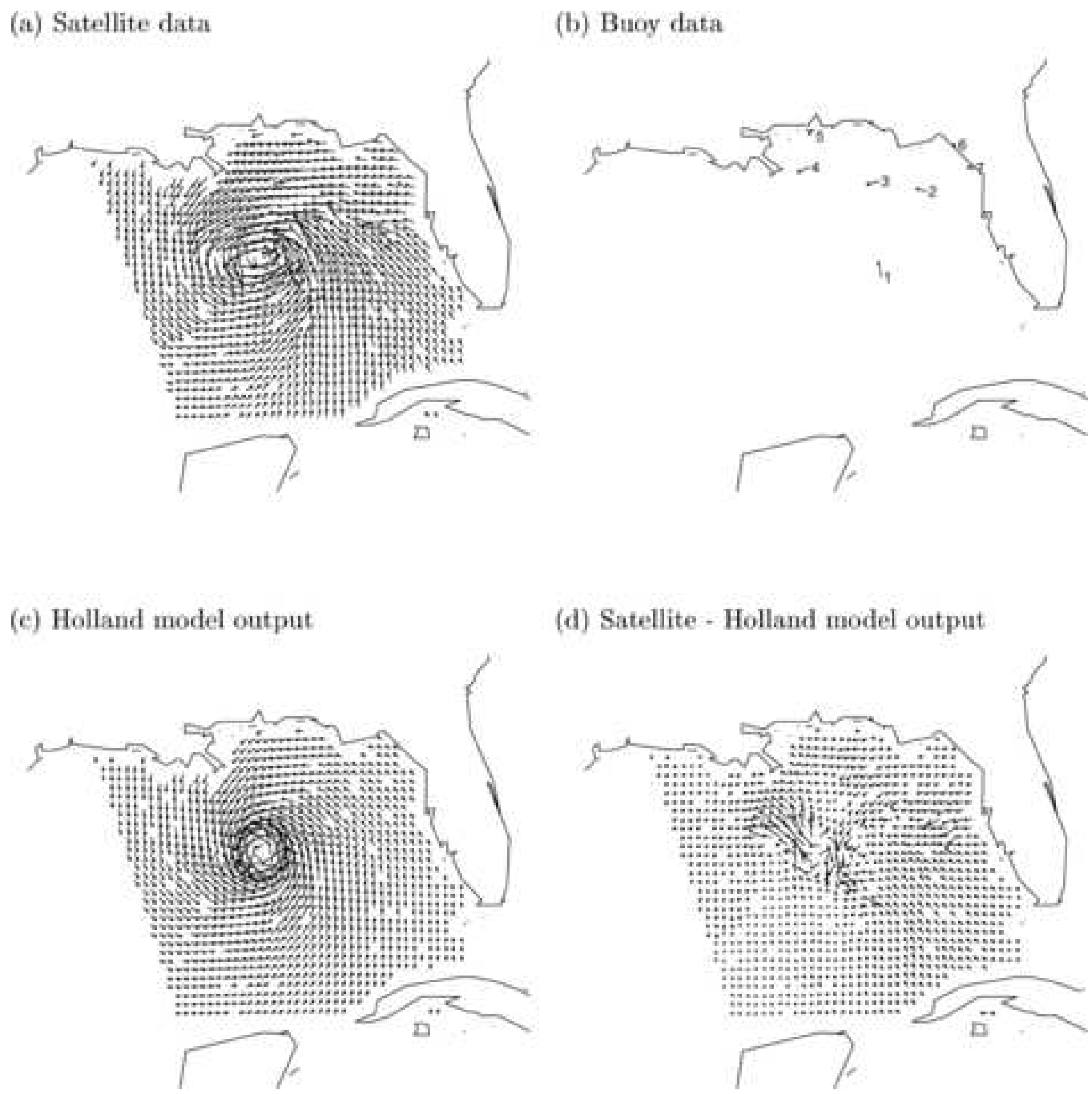

FIG. 1. *Plot of various types of wind field data/output for Hurricane Ivan on September 15, 2004.*

$B = 1.9$ using the meteorological data from the national hurricane center (http://www.nhc.noaa.gov) and recommendations of Hsu and Yan (1998). The output from this model for Hurricane Ivan is plotted in Figure 1(c). By construction, the Holland model output is symmetric with respect to the storm's center, which does not agree with the satellite observations in Figure 1(a).

**3. The spatial stick-breaking (SSB) prior.** In this section we develop a univariate semiparametric spatial model for data from a single source. The spatial stick-breaking prior developed here is incorporated into our model for the wind fields data in Section 4. Let $y(\mathbf{s})$, the observed value at site

$\mathbf{s} = (s_1, s_2)$, have the model

$$y(\mathbf{s}) = \mu(\mathbf{s}) + x(\mathbf{s})'\boldsymbol{\beta} + \varepsilon(\mathbf{s}), \tag{4}$$

where $\mu(\mathbf{s})$ is a spatial random effect, $x(\mathbf{s})$ is a vector of covariates for site $\mathbf{s}$, $\boldsymbol{\beta}$ are the regression parameters and $\varepsilon(\mathbf{s}) \overset{\text{i.i.d.}}{\sim} N(0, \sigma^2)$.

The spatial effects are each assigned a different prior distribution, that is, $\mu(\mathbf{s}) \sim F(\mathbf{s})$. The distributions $F(\mathbf{s})$ are unknown and smoothed spatially. Extending (1) to depend on $\mathbf{s}$, the prior for $F(\mathbf{s})$ is the potentially infinite mixture

$$F(\mathbf{s}) \overset{d}{=} \sum_{i=1}^{m} p_i(\mathbf{s})\delta(\theta_i), \tag{5}$$

where $p_1(\mathbf{s}) = V_1(\mathbf{s})$, $p_i(\mathbf{s}) = V_i(\mathbf{s}) \prod_{j=1}^{i-1}(1 - V_j(\mathbf{s}))$ for $i > 1$, and $V_i(\mathbf{s}) = w_i(\mathbf{s})V_i$. The distributions $F(\mathbf{s})$ are related through their dependence on the $V_i$ and $\theta_i$, which are given the priors $V_i \sim \text{Beta}(a, b)$ and $\theta_i \sim N(0, \tau^2)$, each independent across $i$. However, the distributions vary spatially according to the kernel functions $w_i(\mathbf{s})$, which are restricted to the interval $[0, 1]$. The function $w_i(\mathbf{s})$ is centered at knot $\boldsymbol{\psi}_i = (\psi_{1i}, \psi_{2i})$ and the spread is controlled by the bandwidth parameter $\boldsymbol{\epsilon}_i = (\epsilon_{1i}, \epsilon_{2i})$. Both the knots and the bandwidths are modeled as unknown parameters with priors that are independent of the $V_i$ and $\theta_i$. The knots $\boldsymbol{\psi}_i$ are given independent uniform priors over the bounded spatial domain (this is generalized in Section 4). The bandwidths can be modeled as equal for each kernel function or varying across kernel functions following prior distributions.

Although there are many possible kernel functions, Table 1 gives two examples. Uniform kernels offer bounded support. This is an attractive feature when modeling hurricane wind fields because wind behavior may be different in different subregions, for example, in the center of the storm versus the periphery. We compare uniform kernels with squared-exponential kernels. Squared-exponential kernels decay slowly in space which may be desirable in other applications.

To ensure that the stick-breaking prior is proper, we must choose priors for $\epsilon_i$ and $V_i$ so that $\sum_{i=i}^{m} p_i(\mathbf{s}) = 1$ almost surely for all $\mathbf{s}$. Appendix A.1 shows that the SSB prior with infinite $m$ is proper if $E(V_i) = a/(a + b)$ and $E[w_i(\mathbf{s})]$ [where the expectation is taken over $(\boldsymbol{\psi}_i, \boldsymbol{\epsilon}_i)$] are both positive. For finite $m$, we can ensure that $\sum_{i=i}^{m} p_i(\mathbf{s}) = 1$ for all $\mathbf{s}$ by setting $V_m(\mathbf{s}) \equiv 1$ for all $\mathbf{s}$. This is equivalent to truncating the infinite mixture by attributing all of the mass from the terms with $i \geq m$ to $p_m(\mathbf{s})$.

In practice, allowing $m$ to be infinite is often unnecessary and computationally infeasible. Choosing the number of components in a mixture model is notoriously problematic. Fortunately, in this setting the truncation error can easily be accessed by inspecting the distribution of $p_m(\mathbf{s})$, the mass of the final component of the mixture. The number of components $m$ can be chosen by generating samples from the prior distribution of $p_m(\mathbf{s})$. We increase $m$ until $p_m(\mathbf{s})$ is satisfactorily small for each site $\mathbf{s}$. Also, the truncation error is monitored by inspecting the posterior distribution of $p_m(\mathbf{s})$, which is readily available from the MCMC samples.

TABLE 1
*Examples of kernel functions and the induced functions $\gamma(\mathbf{s},\mathbf{s}')$, where $h_1 = |s_1 - s_1'| + |s_2 - s_2'|$, $h_2 = \sqrt{(s_1 - s_1')^2 + (s_2 - s_2')^2}$, $I(\cdot)$ is the indicator function, and $x^+ = \max(x,0)$*

| **Name** | $w_i(\mathbf{s})$ | **Model for** $\epsilon_{1i}$ **and** $\epsilon_{2i}$ | $\gamma(\mathbf{s},\mathbf{s}')$ |
|---|---|---|---|
| Uniform | $\prod_{j=1}^2 I(\lvert s_j - \psi_{ji}\rvert < \frac{\epsilon_{ji}}{2})$ | $\epsilon_{1i}, \epsilon_{2i} \equiv \lambda$ | $\prod_{j=1}^2 (1 - \frac{\lvert s_j - s_j'\rvert}{\lambda})^+$ |
| Uniform | $\prod_{j=1}^2 I(\lvert s_j - \psi_{ji}\rvert < \frac{\epsilon_{ji}}{2})$ | $\epsilon_{1i}, \epsilon_{2i} \sim \mathrm{Expo}(\lambda)$ | $\exp(-h_1/\lambda)$ |
| Squared exp. | $\prod_{j=1}^2 \exp(-\frac{(s_j-\psi_{ji})^2}{\epsilon_{ji}^2})$ | $\epsilon_{1i}, \epsilon_{2i} \equiv \lambda^2/2$ | $0.5\exp(-\frac{h_2^2}{\lambda^2})$ |
| Squared exp. | $\prod_{j=1}^2 \exp(-\frac{(s_j-\psi_{ji})^2}{\epsilon_{ji}^2})$ | $\epsilon_{1i}, \epsilon_{2i} \sim \mathrm{IG}(1.5, \frac{\lambda^2}{2})$ | $0.5/(1 + (\frac{h_2}{\lambda})^2)$ |

Assuming finite $m$, the spatial stick-breaking model can be written as a mixture model where $g(\mathbf{s}) \in \{1,\ldots,m\}$ indicates site $\mathbf{s}$'s group, that is,

$$
\begin{aligned}
y(\mathbf{s}) &= \theta_{g(s)} + \mathbf{x}(\mathbf{s})'\boldsymbol{\beta} + \varepsilon(\mathbf{s}), \qquad \text{where } \varepsilon(\mathbf{s}) \overset{\text{i.i.d.}}{\sim} N(0,\sigma^2), \\
\theta_j &\overset{\text{i.i.d.}}{\sim} N(0,\tau^2), \qquad j = 1,\ldots,m, \\
g(\mathbf{s}) &\sim \mathrm{Categorical}(p_1(\mathbf{s}),\ldots,p_m(\mathbf{s})), \\
p_j(\mathbf{s}) &= w_j(\mathbf{s})V_j \prod_{k<j}[1 - w_k(\mathbf{s})V_k],
\end{aligned} \tag{6}
$$

where $\mu(\mathbf{s}) = \theta_{g(s)}$, $V_j \overset{\text{i.i.d.}}{\sim} \mathrm{Beta}(a,b)$, and $\prod_{k<j}[1-w_k(\mathbf{s})V_k] = 1$ for $j = 1$. To complete the Bayesian model, we specify priors for the hyperparameters. The regression parameters $\boldsymbol{\beta}$ can be given vague normal priors. In the analysis of Hurricane Ivan in Section 5, the mean term $\mathbf{x}(\mathbf{s})'\boldsymbol{\beta}$ is replaced by the Holland model output. The parameters that control the beta prior for the $V_i$, $a$ and $b$, have independent Uniform$(0,10)$ priors, and the variances $\sigma^2$ and $\tau^2$ have InvGamma$(0.01,0.01)$ priors. We also tried InvGamma$(0.5,0.005)$ priors for the variances and found that the prior had little effect. The knots that control the center of the kernel functions, $\boldsymbol{\psi}_j$, are given uniform priors over the spatial domain and examples of priors for bandwidth parameters $\boldsymbol{\epsilon}_j$ are given in Table 1. The prior for the bandwidths depend on a range parameter, $\lambda$, which is given a Uniform$(0,\lambda_{\max})$ prior. We take $\lambda_{\max}$ to be the maximum distance between any pair of points in the spatial grid. This model can be implemented using WinBUGS. WinBUGS can be freely downloaded from http://www.mrc-bsu.cam.ac.uk/bugs/.

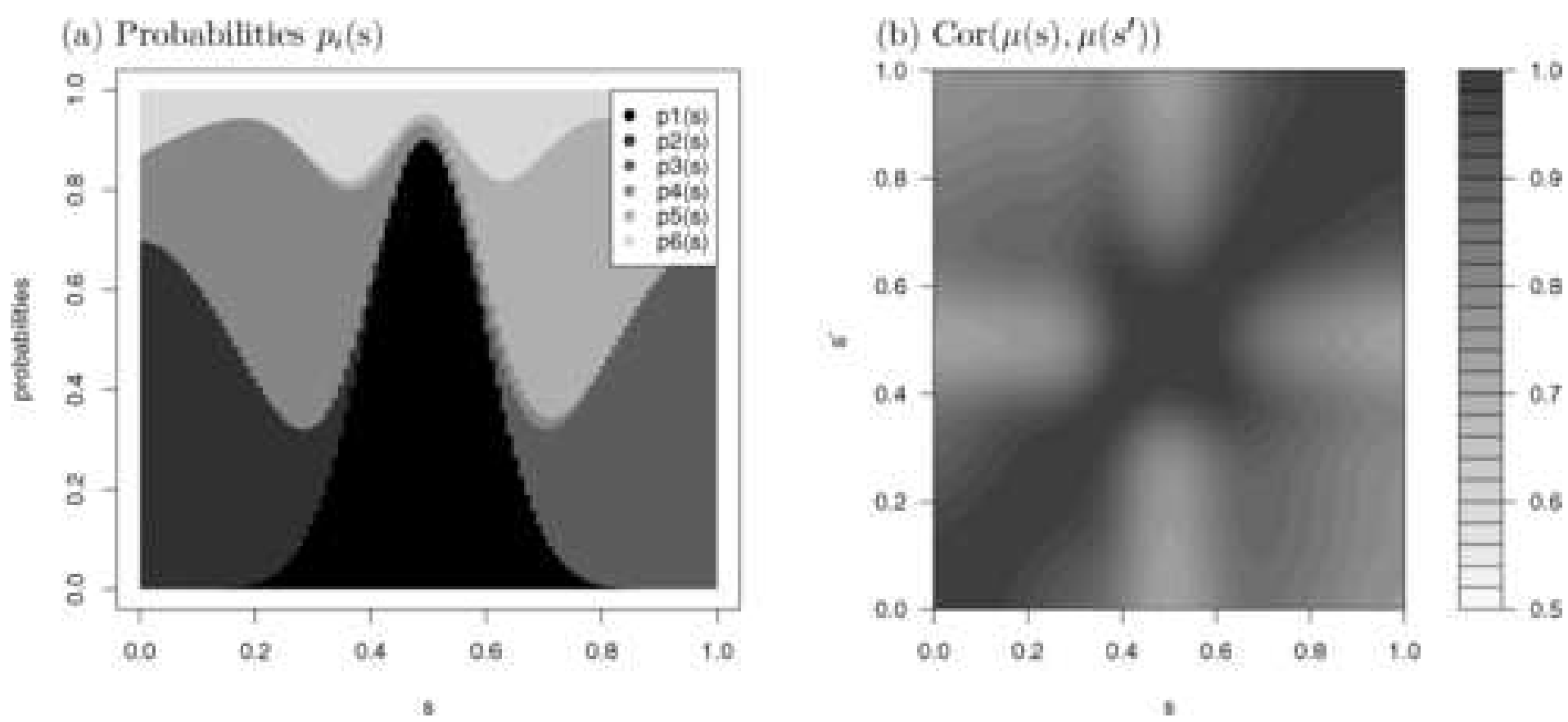

FIG. 2. *Example to illustrate the spatial stick-breaking prior. In this example, the spatial domain is the one-dimensional interval* $(0,1)$ *and the model has Gaussian kernels with knots* $\psi = (0.5, 0.0, 1.0, 0.2, 0.8)$, *bandwidths* $\epsilon = (0.1, 0.2, 0.2, 0.2, 0.2)$ *and* $V = (0.9, 0.7, 0.7, 0.9, 0.9)$. *Panel* (a) *shows the masses* $p_i(\mathbf{s})$ *and panel* (b) *shows the correlation between* $\mu(\mathbf{s})$ *and* $\mu(\mathbf{s}')$.

The mixture model in (6) is nowhere continuous unless uniform kernels are selected and $V_i \in \{0, 1\}$ for all $i$. An alternative suggested by a referee is

$$g(s) = j \qquad \text{where } p_j(s) = \max\{p_1(s), \ldots, p_m(s)\}. \tag{7}$$

This would result in a piece-wise constant random tessellation model which may be preferred for smooth spatial data. However, to avoid oversmoothing micro-scale phenomena in hurricanes, we use the everywhere discontinuous model in (6).

Figure 2(a) illustrates the spatially varying weights of the stick-breaking prior for a one-dimensional example with $m = 6$ and squared exponential kernel functions. We arbitrarily select knots $\psi = (0.5, 0.0, 1.0, 0.2, 0.8)$, bandwidths $\epsilon = (0.1, 0.2, 0.2, 0.2, 0.2)$ and $V = (0.9, 0.7, 0.7, 0.9, 0.9)$. The first kernel function is centered at $s = 0.5$. Since $V_1 = 0.9$, the mass for the first component for $s = 0.5$ is $p_1(0.5) = 0.9$ and decreases as $s$ moves away from 0.5. The second and third kernel functions are centered at $s = 0.0$ and $s = 1.0$ respectively, and dominate the probabilities near the edges. For this example, $p_m(\mathbf{s})$ is as large as 0.2, suggesting $m$ should be increased to give an acceptable approximation to the infinite spatial stick-breaking prior.

Understanding the spatial correlation function is crucial for analyzing spatial data. Although the spatial stick-breaking prior forgoes the Gaussian assumption for the spatial random effects, we can still compute and investigate the covariance function. Conditional on the probabilities $p_j(\mathbf{s})$ (but not the locations $\theta_j$), the covariance between two observations is

$$\operatorname{cov}(y(\mathbf{s}), y(\mathbf{s}')) = \tau^2 P(\mu(\mathbf{s}) = \mu(\mathbf{s}')) = \tau^2 \sum_{j=1}^{m} p_j(\mathbf{s}) p_j(\mathbf{s}'). \tag{8}$$

Figure 2(b) maps the correlation function induced by the probabilities in Figure 2(a). For these probabilities, the correlation is not simply a function of distance between points, that is, the correlation is nonstationary. For example, the correlation is near one for all sites in $(0.4, 0.6)$ due to the large probability for the first component throughout the region. In contrast, the correlation between nearby sites is smaller near $s = 0.35$ and $s = 0.65$ where several components have substantial probability.

As shown in Appendix A.2, integrating over $(V_i, \boldsymbol{\psi}_i, \boldsymbol{\epsilon}_i)$ and letting $m \to \infty$ gives

$$\operatorname{Var}(y(\mathbf{s})) = \sigma^2 + \tau^2, \tag{9}$$

$$\operatorname{Cov}(y(\mathbf{s}), y(\mathbf{s}')) = \tau^2 \gamma(\mathbf{s}, \mathbf{s}') \left[ 2 \frac{a+b+1}{a+1} - \gamma(\mathbf{s}, \mathbf{s}') \right]^{-1}, \tag{10}$$

where

$$\gamma(\mathbf{s}, \mathbf{s}') = \frac{\int\int w_i(\mathbf{s}) w_i(\mathbf{s}') p(\boldsymbol{\psi}_i, \boldsymbol{\epsilon}_i)\, d\boldsymbol{\psi}_i\, d\boldsymbol{\epsilon}_i}{\int\int w_i(\mathbf{s}) p(\boldsymbol{\psi}_i, \boldsymbol{\epsilon}_i)\, d\boldsymbol{\psi}_i\, d\boldsymbol{\epsilon}_i} \in [0, 1]. \tag{11}$$

Since $(V_i, \boldsymbol{\psi}_i, \boldsymbol{\epsilon}_i)$ have independent priors that are uniform over the spatial domain, integrating over these parameters gives a stationary prior covariance. However, Figure 2 illustrates that the conditional covariance can be nonstationary. Therefore, we conjecture the spatial stick-breaking model is more robust to nonstationarity than traditional stationary Kriging methods.

If $b/(a+1)$ is large, that is, the $V_i$ are generally small and there are many terms in the mixture with significant mass, the correlation between $y(\mathbf{s})$ and $y(\mathbf{s}')$ is approximately propotional to $\gamma(\mathbf{s}, \mathbf{s}')$. Table 1 gives the function $\gamma(\mathbf{s}, \mathbf{s}')$ for several examples of kernel functions and shows that different kernels can produce very different correlation functions. For example, $\gamma(\mathbf{s}, \mathbf{s}')$ under the uniform kernel with exponential priors for the bandwidth parameters is the familiar exponential correlation function. If the bandwidths are shared across kernel functions, $\gamma(\mathbf{s}, \mathbf{s}')$ is proportional to a squared exponential covariance under squared exponential kernel functions. The uniform kernel with common bandwidth parameter $\lambda$ gives compact support, as observations separated by more than $\lambda$ spatial units are uncorrelated.

**4. A multivariate spatial model for wind fields data.** Let $u(\mathbf{s})$ and $v(\mathbf{s})$ be the underlying wind speed in the west/east and north/south directions, respectively, for spatial location $\mathbf{s}$. As described in Section 2, there are two types of observed wind data: $u_T(\mathbf{s})$ and $v_T(\mathbf{s})$ are satellite measurements and $u_B(\mathbf{s})$ and $v_B(\mathbf{s})$ are buoy measurements. Our model for these data is

$$\begin{aligned} u_T(\mathbf{s}) &= a_u + u(\mathbf{s}) + e_{uT}(\mathbf{s}), \qquad & v_T(\mathbf{s}) &= a_v + v(\mathbf{s}) + e_{vT}(\mathbf{s}), \\ u_B(\mathbf{s}) &= u(\mathbf{s}) + e_{uB}(\mathbf{s}), & v_B(\mathbf{s}) &= v(\mathbf{s}) + e_{vB}(\mathbf{s}), \end{aligned} \tag{12}$$

where $\{e_{uT}, e_{vT}, e_{uB}, e_{vB}\}$ are independent (with each other and with the underlying winds), zero mean, Gaussian errors, each with its own variance, and $\{a_u, a_v\}$ account for additive bias in the satellite data. Of course, the buoy data may also have bias, but it is impossible to identify bias from both sources, so we attribute all the bias to the satellite measurements. It is also possible to add multiplicative bias terms, but with the small number of buoy observations it will be difficult to identify both types of bias and Foley and Fuentes (2006) found that the primary source of bias is additive.

The underlying orthogonal wind components $u(\mathbf{s})$ and $v(\mathbf{s})$ are modeled as a mixture of a deterministic wind model and a semiparametric multivariate spatial process

$$\begin{aligned} u(\mathbf{s}) &= H_u(\mathbf{s}) + R_u(\mathbf{s}), \\ v(\mathbf{s}) &= H_v(\mathbf{s}) + R_v(\mathbf{s}), \end{aligned} \tag{13}$$

where $H_u(\mathbf{s})$ and $H_v(\mathbf{s})$ are the orthogonal components of the deterministic Holland model in (3) and $\mathbf{R}(\mathbf{s}) = (R_u(\mathbf{s}), R_v(\mathbf{s}))'$ follows a multivariate extension of the non-Gaussian spatial stick-breaking prior of Section 3. We take $\mathbf{R}(\mathbf{s}) \sim F(\mathbf{s})$, where $F$ has the stick-breaking prior in (5) modified so the two-dimensional locations $\boldsymbol{\theta}_i$ have multivariate normal priors $\boldsymbol{\theta}_i \stackrel{\text{i.i.d.}}{\sim} N(0, \Sigma)$, where $\Sigma$ is a $2 \times 2$ covariance matrix that controls the association between the two wind components. The covariance $\Sigma$ has an $\mathrm{InvWish}(0.1, 0.1 I_2)$ prior and after transforming the spatial grid to be contained in the unit square, the spatial knots $\psi_{s_1 i}$ and $\psi_{s_2 i}$ have independent Beta$(1.5, 1.5)$ priors to encourage knots to lie near the center of the hurricane where the wind is most volatile. Also, we take the spatial range $\lambda \sim \mathrm{Uniform}(0, 1)$.

Assuming the same priors for the $p_i(\mathbf{s})$ as in Section 3 and following the same steps as in Appendix A.2, it can be shown that $\mathrm{Cov}(\mathbf{R}(\mathbf{s}), \mathbf{R}(\mathbf{s}'))$ is separable, that is, the product of the spatial covariate and the cross-dependency covariance matrix $\Sigma$. This could be generalized by allowing the prior covariance of the $\boldsymbol{\theta}_i$ to vary spatially. Alternatively, the spatial stick-breaking prior could be combined with the linear model of coregionalization to give a nonseparable multivariate spatial model by modeling the $u$ and $v$ components of $\mathbf{R}(\mathbf{s})$ as linear combinations of univariate spatial terms given spatial stick-breaking priors described in Section 3.

**5. Analysis of Hurricane Ivan's wind field.** We fit three models to 182 satellite observations and 7 buoy observations for the Hurricane Ivan. We use the multivariate SSB model in Section 4 with both uniform and squared-exponential kernels. Also, to illustrate the effect of relaxing the normality assumption, we also fit a fully-Gaussian Bayesian Kriging model [Handcock and Stein (1993)] that replaces the stick-breaking prior for $\mathbf{R}(\mathbf{s}) = (R_s(\mathbf{s}), R_v(\mathbf{s}))'$ in (13) with a zero-mean Gaussian prior with separable covariance

$$\mathrm{Var}(\mathbf{R}(\mathbf{s})) = \Sigma \quad \text{and} \quad \mathrm{Cov}(\mathbf{R}(\mathbf{s}), \mathbf{R}(\mathbf{s}')) = \Sigma \times \exp(-\|\mathbf{s} - \mathbf{s}'\|/\lambda), \tag{14}$$

where $\Sigma$ controls the dependency between the wind components at a given location and $\lambda$ is a spatial range parameter. The covariance parameters $\Sigma$ and $\lambda$ have the same priors as the covariance parameters in Section 4.

Since our primary objective is to predict wind vectors at unmeasured locations to use as inputs for numerical ocean models, we compare models in terms of expected mean squared prediction error [Laud and Ibrahim (1995) and Gelfand and Ghosh (1998)], that is,

$$\begin{aligned} \mathit{EMSPE} = E\Bigg(&\sum_s (u_T(\mathbf{s}) - \tilde{u}_T(\mathbf{s}))^2 + (v_T(\mathbf{s}) - \tilde{v}_T(\mathbf{s}))^2 \\ &+ \sum_s (u_B(\mathbf{s}) - \tilde{u}_B(\mathbf{s}))^2 + (v_B(\mathbf{s}) - \tilde{v}_B(\mathbf{s}))^2\Bigg), \end{aligned} \tag{15}$$

where, say, $\tilde{u}_T(\mathbf{s})$ is viewed as a replicate of the observed $u$ component of the satellite measurement at site $\mathbf{s}$, the summation is taken over all observation locations, and the expectation is taken over the full posterior of all the parameters in the model. This model selection criteria favors predictive models centered near the observed data with small predictive variances.

The EMSPE is smaller for the semiparametric model uniform kernels ($\mathit{EMSPE} = 3.46$) than for the semiparametric model squared exponential kernels ($\mathit{EMSPE} = 4.19$) and the fully-Gaussian model ($\mathit{EMSPE} = 5.17$). Figures 3(a) and 3(b) show that the squared residuals from the fully-Gaussian fit are near zero for most of the spatial domain but are large near the center of the hurricane for both components. The Gaussian model oversmooths in this area with high volatility in the underlying wind surface. In contrast, the semiparametric model with uniform kernel functions is able to capture the peaks near the eye of the hurricane and the squared residuals [Figures 3(c) and 3(d)] show less spatial structure than the residuals from the Gaussian model.

Figure 4 summarizes the posterior from the spatial stick-breaking prior with uniform kernel functions. The fitted values in Figures 4(a) and 4(b) vary rapidly near the center of the storm and are fairly smooth in the periphery.

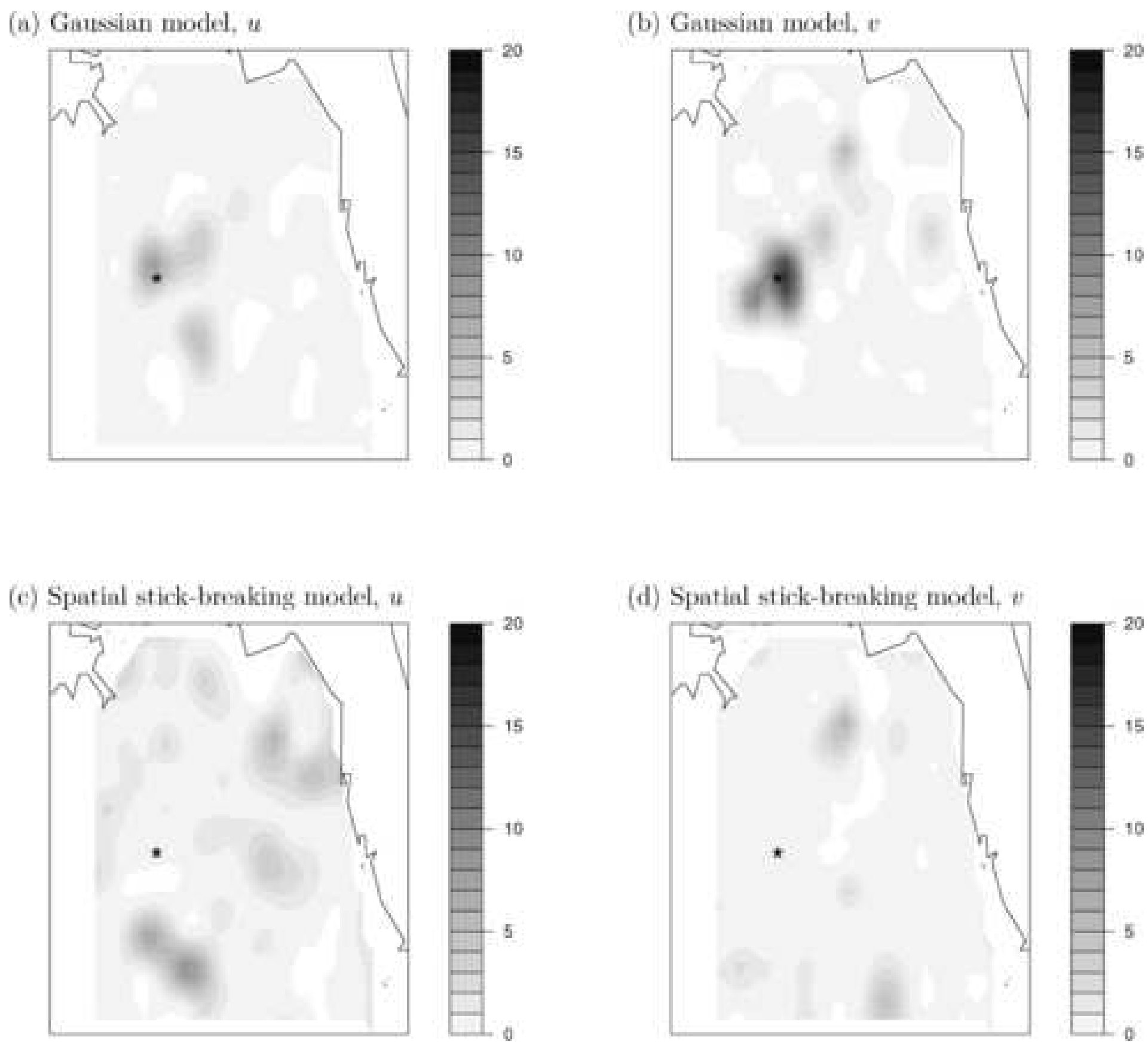

FIG. 3. *Squared residuals (value-posterior mean) for the* $u$ *and* $v$ *components of the Gaussian and spatial stick-breaking model with uniform kernels. The "*$\star$*" represents the storm's center.*

After accounting for the Holland model, the correlation between the residual $u$ and $v$ components $R_u(\mathbf{s})$ and $R_v(\mathbf{s})$ [$\Sigma_{12}/\sqrt{\Sigma_{11}\Sigma_{22}}$, where $\Sigma_{kl}$ is the $(k,l)$ element of $\Sigma$] is generally negative [Figure 4(c)], confirming the need for a multivariate analysis. Figure 4(d) plots the posterior of the parameter that controls the size of bandwidths, $\lambda$. The posterior median of $\lambda$ is 0.17, so on average the uniform kernels span about 17% of the spatial domain.

The satellite data are significantly biased relative to the buoy data. The 95% posterior intervals for the bias terms $a_u$ and $a_v$ are $(-6.91, -2.16)$ and $(0.04, 4.38)$ respectively. The biases seem to be driven by the third buoy's wind vector in Figure 1(b), which is quite different from the nearby satellite observations in Figure 1(a).

To show that the semiparametric model with uniform kernel functions fits the data well, we randomly (across $u$ and $v$ components and buoy and satellite data) set aside 10% of the observations and compute 95% predictive intervals for the missing observations. The prediction intervals contain 94.7% (18/19) of the deleted $u$ components and 95.2% (20/21) of the deleted $v$ components. These statistics suggest that our model is well calibrated.

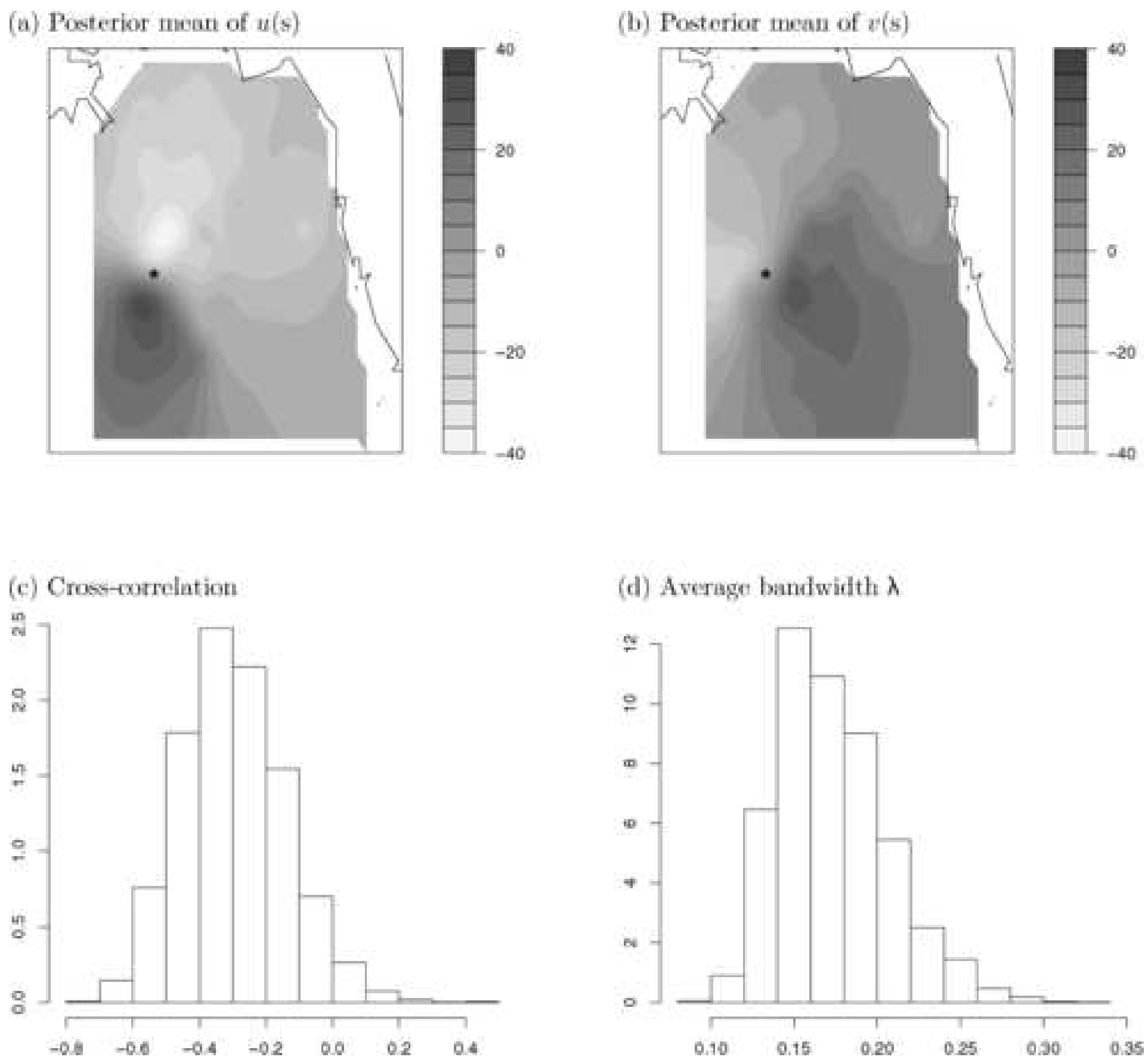

FIG. 4. *Summary of the posterior of the spatial stick-breaking model with uniform kernels. Panels* (a) *and* (b) *give the posterior mean surface for the u and v components, panel* (c) *shows the posterior of the cross-correlation between the residual wind components* $R_u(\mathbf{s})$ *and* $R_v(\mathbf{s})$ *(*$\Sigma_{12}/\sqrt{\Sigma_{11}\Sigma_{22}}$*), and panel* (d) *plots the posterior of the parameter that controls the average kernel bandwidth* $\lambda$ *assuming the spatial grid has been transformed to lie in the unit square. The "*$\star$*" represents the storm's center.*

**6. Discussion.** Modeling hurricane wind fields is an important and challenging problem. This paper presents a semiparametric multivariate spatial model for these data. The semiparametric model avoids oversmoothing near the center of Hurricane Ivan's wind field, resulting in a well-calibrated predictive model. Gaussian models with highly-structured covariance functions, for example, Wikle et al. (2001) and Fuentes et al. (2005), are an alternative. However, our non-parametric model offers greater flexibility by allowing for nonstationarity and nonnormality which is advantageous when building an automated procedure.

In the statistical model for wind fields data, the spatial random effects with the spatial stick-breaking prior are mixed with independent normal errors. An extension of this model would be to replace the independent normal effects with a Gaussian spatial process. This would give a mixture of two spatial terms: a semiparametric term to handle discontinuities and a Gaussian process which performs well in smooth areas. A mixture of this nature has been considered by Lawson and Clark (2002), who propose a fully-parametric mixture of spatial models for disease mapping with a real spatial data. As Lawson and Clark point out, it can be difficult to identify the contribution of each component of the mixture, but using a combination of spatial terms can lead to an improvement in fit.

This paper focused on estimating the wind field at a single time point because satellite data are only available twice daily. However, the spatial stick-breaking prior developed here could be extended to the spatiotemporal setting to improve real-time estimates. One possibility is to use three-dimensional kernel functions in space and time. An alternative spatiotemporal model would be an extension of the dynamic linear model of Gelfand, Banerjee and Gammerman (2005b), that is, $\mathbf{R}(\mathbf{s},t) = B\mathbf{R}(\mathbf{s},t-1) + \Delta(\mathbf{s},t)$, where $\mathbf{R}(\mathbf{s},t)$ is the vector of residual wind components at location $\mathbf{s}$ at time $t$, $B$ is diagonal with $B_{ii} \in [-1,1]$, and $\Delta(\mathbf{s},t)$ is the vector of changes from time $t-1$ to time $t$. The spatial stick-breaking prior could be applied to the mean at the first time point, $\mathbf{R}(\mathbf{s},1)$, and each $\Delta(\mathbf{s},t)$.

## APPENDIX A.1: PROPRIETY OF THE SSB PRIOR

For infinite $m$, Ishwaran and James (2001) show that $\sum_{i=i}^{m} p_i(\mathbf{s}) = 1$ almost surely if and only if $\sum_{i=1}^{\infty} E(\log(1-V_i(\mathbf{s}))) = -\infty$. Applying Jensen's inequality,

$$E[\log(1-V_i(\mathbf{s}))] \le \log[E(1-V_i(\mathbf{s}))] = \log[1-E\{w_i(\mathbf{s})\}E(V_i)].$$

If both $E\{w_i(\mathbf{s})\}$ and $E(V_i)$ are positive, $\log(1-E\{w_i(\mathbf{s})\}E(V_i))$ is negative and

$$\sum_{i=1}^{\infty} E[\log(1-V_i(\mathbf{s}))] \le \sum_{i=1}^{\infty} \log(1-E\{w_i(\mathbf{s})\}E(V_i)) = -\infty.$$

## APPENDIX A.2: COV($\mu(\mathbf{S}), \mu(\mathbf{S}')$)

Due to the discrete nature of the stick-breaking prior, $\mathrm{Cov}(\mu(\mathbf{s}),\mu(\mathbf{s}')) = \tau^2\,\mathrm{Prob}(\mu(\mathbf{s}) = \mu(\mathbf{s}'))$:

$$\mathrm{Prob}[\mu(\mathbf{s}) = \mu(\mathbf{s}')|V_i, \boldsymbol{\psi}_i, \boldsymbol{\epsilon}_i]$$

$$= \sum_{i=1}^{\infty} p_i(\mathbf{s}) p_i(\mathbf{s}')$$
$$= \sum_{i=1}^{\infty} \left[ w_i(\mathbf{s}) w_i(\mathbf{s}') V_i^2 \prod_{j<i} (1 - (w_j(\mathbf{s}) + w_j(\mathbf{s}'))V_j + w_j(\mathbf{s}) w_j(\mathbf{s}') V_j^2) \right].$$

Integrating over the $(V_i, \boldsymbol{\psi}_i, \boldsymbol{\epsilon}_i)$ gives

$$\mathrm{Prob}(\mu(\mathbf{s}) = \mu(b\mathbf{s}')) = c_2 \tilde{v}_2 \sum_{i=1}^{\infty} [1 - 2c_1 \tilde{v}_1 + c_2 \tilde{v}_2]^{i-1},$$

where $c_1 = \int\int w_i(\mathbf{s}) p(\boldsymbol{\psi}_i, \boldsymbol{\epsilon}_i)\, d\boldsymbol{\psi}_i\, d\boldsymbol{\epsilon}_i$, $c_2 = \int\int w_i(\mathbf{s}) w_i(\mathbf{s}') p(\boldsymbol{\psi}_i, \boldsymbol{\epsilon}_i)\, d\boldsymbol{\psi}_i\, d\boldsymbol{\epsilon}_i$, $\tilde{v}_1 = E(V_1) = a/(a+b)$, and $\tilde{v}_2 = E(V_1^2) = a(a+1)/[(a+b)(a+b+1)]$. Since $1 - 2c_1\tilde{v}_1 + c_2\tilde{v}_2 = E[(1 - p_i(\mathbf{s}))(1 - p_i(\mathbf{s}'))] \in [0, 1]$, we apply the formula for the sum of a geometric series and simplify, leaving

$$\mathrm{Prob}(\mu(\mathbf{s}) = \mu(\mathbf{s}')) = \frac{c_2 \tilde{v}_2}{2c_1 \tilde{v}_1 - c_2 \tilde{v}_2} = \frac{\gamma(\mathbf{s}, \mathbf{s}')}{2(1 + b/(a+1)) - \gamma(\mathbf{s}, \mathbf{s}')},$$

where $\gamma(\mathbf{s}, \mathbf{s}') = c_2/c_1$.

**Acknowledgments.** The authors thank Professor Lian Xie of the Coastal Fluid Dynamics Laboratory at North Carolina State University and Dr. Kristen Foley of the US EPA for their helpful insight.

DEPARTMENT OF STATISTICS
NORTH CAROLINA STATE UNIVERSITY
2501 FOUNDERS DRIVE
BOX 8203
RALEIGH, NORTH CAROLINA 27695
USA
E-MAIL: reich@stat.ncsu.edu
fuentes@stat.ncsu.edu